\documentclass[reprint,
aps,pra,
10pt,
groupedaddress,
amsmath,amssymb,
floatfix
]{revtex4-1} 

\usepackage{graphicx}
\usepackage{amsmath,amssymb}
\usepackage{mathtools}
\usepackage{dcolumn}
\usepackage{esvect}
\usepackage{booktabs}
\usepackage{titlesec}
\usepackage{siunitx}
\usepackage[svgnames]{xcolor}
\usepackage[colorlinks=true,
linkcolor=MediumBlue,
urlcolor=MediumBlue,
citecolor=MediumBlue]{hyperref}

\titlespacing\section{0pt}{15pt plus 2pt minus 2pt}{6pt plus 2pt minus 2pt}
\titlespacing\subsection{0pt}{15pt plus 2pt minus 2pt}{6pt plus 2pt minus 2pt}

\renewcommand{\vec}[1]{\ensuremath{\mathbf{#1}}}

\begin{document}

\spaceskip=0.75\fontdimen2\font plus 1.0\fontdimen3\font minus 1.0\fontdimen4{0.91}

\title{Single-beam dielectric-microsphere trapping with optical heterodyne detection}

\author{Alexander D. Rider}
\author{Charles P. Blakemore}
\email{cblakemo@stanford.edu}
\author{Giorgio Gratta}
\affiliation{Department of Physics, Stanford University, Stanford, California 94305, USA}

\author{David C. Moore}
\affiliation{Department of Physics, Yale University, New Haven, Connecticut 06520, USA}

\date{\today}

\begin{abstract}
\linespread{1.1}\selectfont
A technique to levitate and measure the three-dimensional position of micrometer-sized dielectric spheres with heterodyne detection is presented. The two radial degrees of freedom are measured by interfering light transmitted through the microsphere with a reference wavefront, while the axial degree of freedom is measured from the phase of the light reflected from the surface of the microsphere. This method pairs the simplicity and accessibility of single-beam optical traps to a measurement of displacement that is intrinsically calibrated by the wavelength of the trapping light and has exceptional immunity to stray light.  A theoretical shot noise limit of $\smash{\SI{1.3e-13}{m/\sqrt{Hz}}}$ for the radial degrees of freedom, and $\smash{\SI{3.0e-15}{m/\sqrt{Hz}}}$ for the axial degree of freedom can be obtained in the system described. The measured acceleration noise in the radial direction is $\smash{\SI{7.5e-5}{(m/s^2)/\sqrt{Hz}}}$. \\

\noindent DOI: \href{https://doi.org/10.1103/PhysRevA.97.013842}{10.1103/PhysRevA.97.013842}
\end{abstract}

\maketitle

\section{INTRODUCTION}

Optical traps for small dielectric particles have been used since the pioneering work of Ashkin and Dziedzic~\cite{Ashkin:1971}. Although many of the initial applications of these traps were in biology and polymer science, where the particles are suspended in a liquid~\cite{Ashkin:1986,Neuman:2004,Bishop:2004,Ether:2015}, trapping and cooling of microspheres (MSs) in a vacuum environment has become a common tool in the fields of optomechanics~\cite{Chang:2010,Li:2011,Li:2013,Yin:2013,Millen:2015,Hoang:2016,Jain:2016,Hempston:2017}, quantum control~\cite{Romero:2010,Kaltenbaek:2012} and fundamental particles and interactions~\cite{Geraci:2008,Geraci:2010,Ether:2015,Moore:2014,Rider:2016}. While several techniques for trapping MSs in vacuum have been proposed and implemented~\cite{Geraci:2008,Li:2011,Gieseler:2012,Asenbaum:2013,Mazilu:2016,Ranjit:2015,Ranjit:2016,Fonseca:2016,Vovrosh:2017}, single-beam traps with an upward propagating, focused laser beam and active feedback have the advantages of simplicity and access to the trapping region to probe the MS.

In the single-beam trap described here, radiation pressure from the beam supports the weight of the MS while recoil against light deflected by the MS provides a restoring force, confining the MS toward the axis of the beam~\cite{Ashkin:1971,Ashkin:1977,Ashkin:1986} where the MS undergoes harmonic motion in three dimensions. Radial (horizontal) feedback forces are applied to the MS by modulating the position of the trap while axial (vertical) feedback forces are applied by modulating the trap beam power. Vacuum operation is required to minimize noise due to collisions between residual gas and the MS. Under vacuum, active feedback is used to stabilize the trap by replacing the damping from residual gas. In this way, the center-of-mass motion of the particle can be damped to obtain \SI{}{mK} effective temperatures~\cite{Li:2011,Gieseler:2012,Asenbaum:2013,Fonseca:2016} with the rest of the system at room temperature.   

The system described here uses heterodyne detection to measure the position of the MS and provide feedback by interfering the light transmitted through, and reflected by, the MS with frequency-shifted phase reference beams. In addition to the simplicity of the single-beam trap, the heterodyne detection technique results in improved immunity to stray sources of light not associated with the MS because only light spatially and temporally coherent with the phase reference beam produces an interference signal at the detector.  The ability to reject scattered light is particularly important for short-distance force sensing applications~\cite{Geraci:2010,Rider:2016}, where objects used to probe the MS may scatter light from the trapping beam, producing background signals.

\section{EXPERIMENTAL SETUP}

The apparatus presented here makes use of \SI{4.8}{\mu m} diameter silica MS~\cite{bangs_laboratories} trapped in vacuum by a single-beam optical trap.  The trapping and phase reference beams are produced by seeding a single frequency, polarization maintaining (PM), Yb-doped fiber amplifier~\cite{nuamp} with light from a $\lambda = \SI{1064}{nm}$, single frequency, distributed feedback, Yb-doped fiber laser~\cite{ethernal}.  The spatial coherence length of this system, $\gg \SI{e3}{m} $, is greater than any differences in optical path length.  The output of the amplifier is passed through a PM fiber splitter to obtain trapping and phase reference beams.  The two branches are sent into fiber-coupled acousto-optic modulators (AOMs) driven at \SI{149.5}{MHz} and \SI{150.0}{MHz} that frequency shift the source light and allow for intensity modulation. Frequency shifting both the trapping light and the phase reference light avoids the amplification of rf signals at the interference frequency which could lead to electronic backgrounds. The reference branch is further split into two branches for use in the axial and radial heterodyne measurements.

\begin{figure*}[t]
\includegraphics[width=2.0\columnwidth]{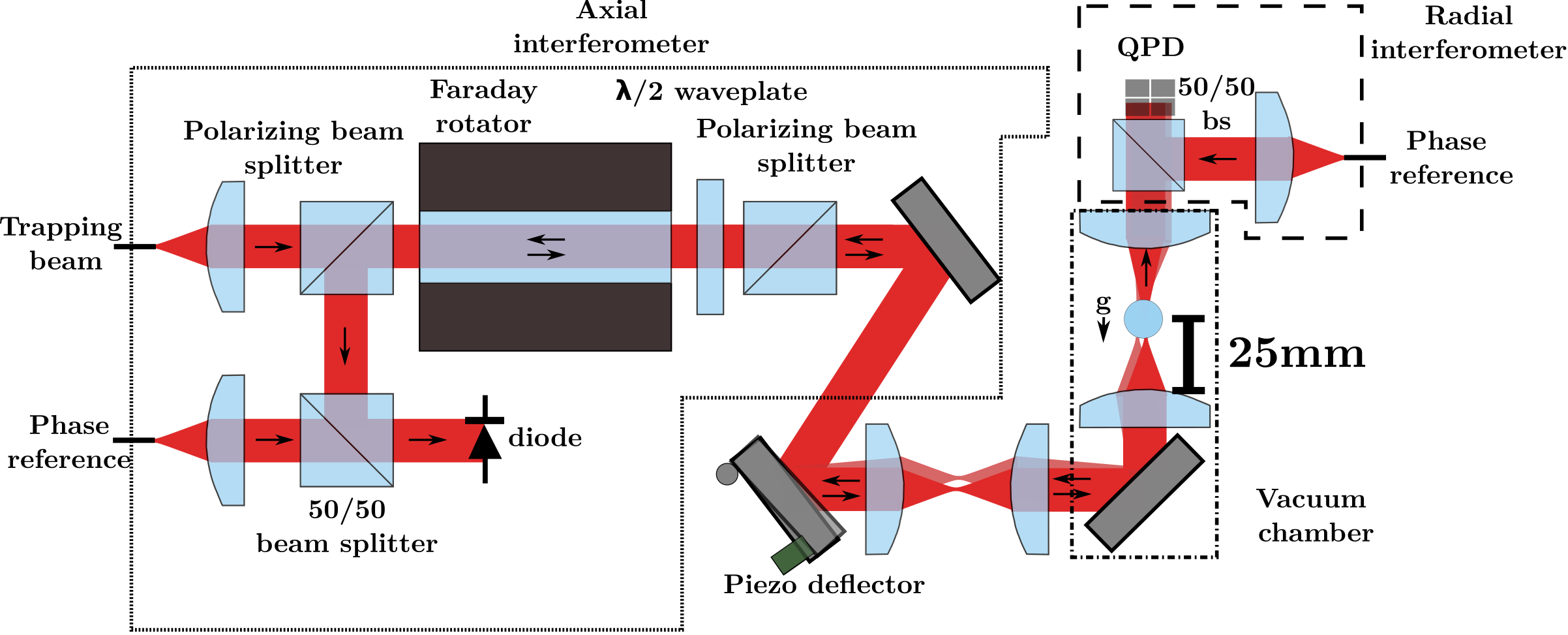}
\caption{Schematic view of the free-space optical system. The output of the fiber carrying the trapping beam is first collimated, then deflected by a high-bandwidth piezo-mounted mirror in the conjugate focal plane of the trap.  This produces translations in the plane of the trap, as indicated by the two closely spaced beams. A telescope is used to adjust the gain of the deflection system. Two identical aspheric lenses inside the vacuum chamber focus the trapping beam and recollimate it. The collimated beam is then recombined with a reference beam on a quadrant photodiode (QPD).  Light that is backscattered by the MS is extracted, recombined with another reference beam and used to interferometrically measure the axial position of the MS.}
\label{fig:freespace_system}
\end{figure*} 

\begin{figure}[b!]  
\includegraphics[width=1.0\columnwidth]{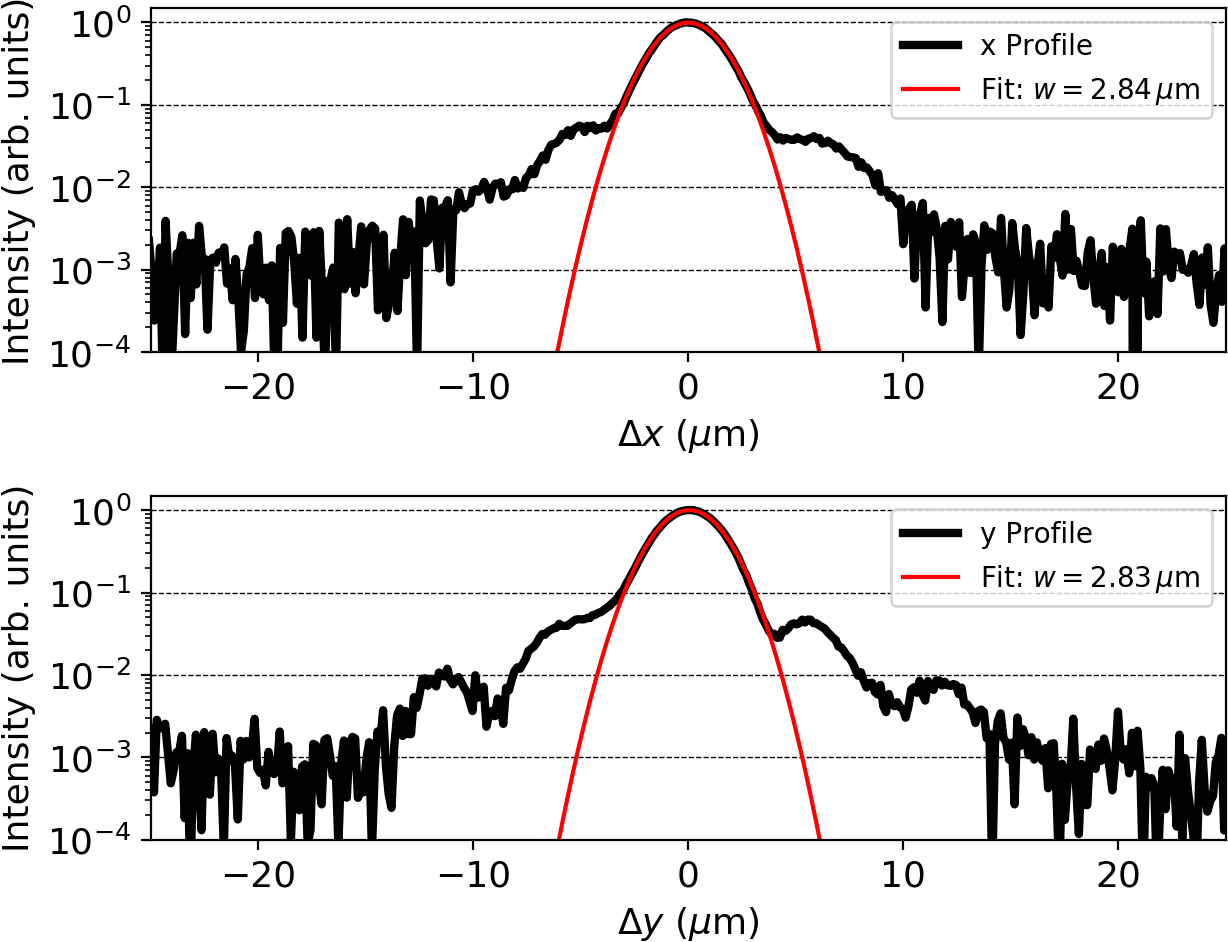}
\caption{Trapping beam profile in the $x$ and $y$ axes at the stable point of the trap.  Fits to a Gaussian profile give $w_{o,x} = \SI{2.84}{\mu m}$ and $w_{o,y} = \SI{2.83}{\mu m}$ (with $w_o$ being the usual Gaussian waist) at the stable point of the trap, which is a small fraction of a Rayleigh range above the focus. Non-Gaussian tails could result from cladding modes within the fiber or imperfections in optical surfaces, as well as small misalignments.}
\label{fig:beam_profile}
\end{figure} 

A simplified schematic of the free space optics forming the trap and providing the position readout for the MS is shown in Fig.~\ref{fig:freespace_system}.  The trapping and reference beams are projected into free space with the fibers providing mode cleaning and flexibility of installation. All optical fiber components are fusion spliced together for reliability. Following the fiber launch, the trapping beam passes through a Faraday isolator which is used to extract the back-propagating light reflected by the MS. The beam is then reflected off a high-bandwidth (\SI{3}{dB} point at \SI{2.5}{kHz}) piezo-actuated deflection mirror imaged into the the Fourier plane of the trap by a telescope. Angling the mirror produces displacements of the trap that are used to apply radial feedback forces to the MS.

The beam is then injected into the vacuum chamber, where it is focused by a \SI{25}{mm} focal length aspheric lens to form the trap. This long free working distance is ideal for many applications requiring access to the trapping region. The beam transmitted through the microsphere is recollimated by an identical aspheric lens, sent out of the vacuum chamber and superposed with a reference beam.  This superposition is projected onto a quadrant photodiode (QPD) from which the radial motion along two axes, $x$ and $y$, is extracted from the interference photocurrents at the difference in modulation frequencies. 

To measure the axial position of the MS, the back-propagating light extracted by the Faraday isolator is interfered with the second phase reference beam. The interference generates an rf photocurrent whose phase encodes the $z$-position, as the path length of the back-propagating light depends on the MS position. This eliminates the need for an auxiliary imaging beam perpendicular to the trapping beam and provides the MS position in absolute units related to the wavelength of the trapping light. 

All relevant optics have been optimized to image the fiber mode into the focal plane of the trap with minimal distortion. A relatively pure fundamental Gaussian spatial mode is important for short-distance force sensing, where devices are brought into close proximity with the MS~\cite{Rider:2016}. The profile of the spatial mode at the stable position in the trap is shown in Fig.~\ref{fig:beam_profile}.  

The axial and the radial signals are first digitized, and then analyzed by a field programmable gate array (FPGA) running the algorithms that generate real-time feedback signals.  In the radial direction, only active damping is applied, so as not to disturb force measurements at frequencies below the trap resonance.

\section{RADIAL DISPLACEMENT CALIBRATION}

While axial displacements are intrinsically calibrated into physical units by the wavelength of the laser, displacements of the radial degree of freedom have to be calibrated empirically. A calibration of the radial position measurement is obtained by measuring the response of the system to known forces at frequencies far below resonance, and then dividing this response by the spring constant of the trap. The response to forces is determined, with active feedback on, by applying an alternating electric field to a MS with a few quanta of charge, as demonstrated in Ref.~\cite{Moore:2014}.  The result of this procedure is $(\Delta V / F)_{\text{meas}} = (7.5 \pm 0.3) \times \SI{e13}{V/N}$ for either radial degree of freedom (DOF), within uncertainties, where $\Delta V$ is the voltage generated in photodetection due to a difference in photocurrent between sides of the QPD and $F$ is the known force applied to the MS. This quantity can then be converted into a calibration constant for position by using $F_{\text{app}} = k x_{\text{MS}}$ where $x_{\text{MS}}$ is the displacement in one of the radial DOFs and the spring constant $k$ is measured by observing the response of the MS to an oscillating electric field of variable frequency. 

It is instructive to compare this empirical calibration constant to the ideal one, calculated from the properties of the system. In principle, this could be derived by solving Mie scattering theory, whereby MS displacements deflect some of the trapping light, and applying simple ray optics.  However, the relationship between the MS displacement, $x_{\text{MS}}$, and the angle by which the light is deflected, $\theta$, can be extracted directly by considering the optical restoring force $F_{\text{opt}}$ for a certain $\theta$.  This force is related to displacements of the MS by the spring constant of the trap $k = m_{\text{MS}} \Omega^2$ where $m_{\text{MS}}$ is the mass of the MS and $\Omega$ is the resonant frequency of the trap. For small displacements causing small $\theta$, $F_{\text{opt}} = (\mathcal{P} / c) \, \theta $, where $\mathcal{P}$ is the power of the beam transmitted through the MS and $c$ is the speed of light. After re-collimation by a lens of focal length $d$, the relationship between the force and the translation of the outgoing beam, $x_B$, is

\vspace*{-0.2cm}
\begin{equation} \label{eq:force_to_displacement}
\frac{F_{\text{opt}}}{x_B} = \frac{\mathcal{P}}{d \, c}.
\end{equation}

The quantity $x_B$ is determined by interfering the beam with a phase reference beam shifted in frequency by an amount $\Delta \omega$, and projecting their superposition onto a segmented photodetector. 

If the transmitted and reference beams are Gaussian with their foci in the detector plane, then $\vec{E}_i(\vec{x}) = \hat{p}_i E_i e^{-\vec{x}^2 / w_i^2}$ with $\hat{p}_i$ being the polarization vectors and $w_i$ being the usual Gaussian waists ($w = 2 \sigma_{\mathcal{P}}$, where $\sigma_{\mathcal{P}}$ the standard deviation of the intensity). Displacing beam ``2'' by $x_B$, and making use of Eq.~(\ref{eq:force_to_displacement}) and $F=kx_{\text{MS}}$, the difference in photocurrent between adjacent segments per MS displacement is given by

\vspace*{-0.2cm}
\begin{equation} \label{eq:current_seg_detector}
\frac{\Delta I}{x_{\text{MS}}} = 4 \xi \sqrt{\frac{\mathcal{P}_1 {\mathcal{P}}_2}{4 \pi}} \frac{w_1^2 }{( w_1^2 + w_2^2)^{3/2}}  \frac{k \, d \, c}{ {\mathcal{P}}_{2} } \text{cos} [ \Delta \omega t + \Delta \phi],
\end{equation}

\noindent where $\xi$ is the responsivity of the photodetector in \SI{}{A/W}, $\mathcal{P}_1$ ($\mathcal{P}_2$)  and $w_1$ ($w_2$) are the power and waist of the phase reference (transmitted) beam, respectively, and beam 2 is displaced. In practice, $\mathcal{P}_1$ can be increased to optimize sensitivity, but $\mathcal{P}_2$ is fixed by the power required to levitate a particular mass of MS. For the parameters of the system described here, $\mathcal{P}_1 = \SI{25}{mW}$, $\mathcal{P}_2 = \SI{1.1}{mW}$, $w_1 = \SI{3.7}{mm}$, $w_2 = \SI{3.0}{mm}$, $k = \SI{2.0e-7}{N/m}$, $\xi = \SI{0.34}{A/W}$, and $d = \SI{25}{mm}$, this corresponds to an ideal difference in photocurrent per MS displacement of \SI{350}{A/m}. A detailed derivation of Eq~(\ref{eq:current_seg_detector}) is included in the appendix.

This value can be compared with the empirically obtained calibration as $\Delta V / F = (\Delta I / x_{\text{MS}}) \cdot (G R_t / k)$, where $G$ is the gain of the readout electronics and $R_t$ is the transimpedance. With $G=259 \pm 3$ set by digital potentiometers, and $R_t = \SI{1}{k\Omega}$, we find that $(\Delta V / F)_{\text{calc}} = \SI{4.5e14}{V/N}$ compared with $(\Delta V / F)_{\text{meas}} = (7.5 \pm 0.3) \times \SI{e13}{V/N}$. We attribute the discrepancy to imperfect mode-matching between the transmitted and the reference beams as well as slight non-Gaussianity of the modes of the two beams, as can be seen in Fig.~\ref{fig:beam_profile}.

\section{DISPLACEMENT NOISE}

\begin{figure*}[ht!]  
\includegraphics[width=2\columnwidth]{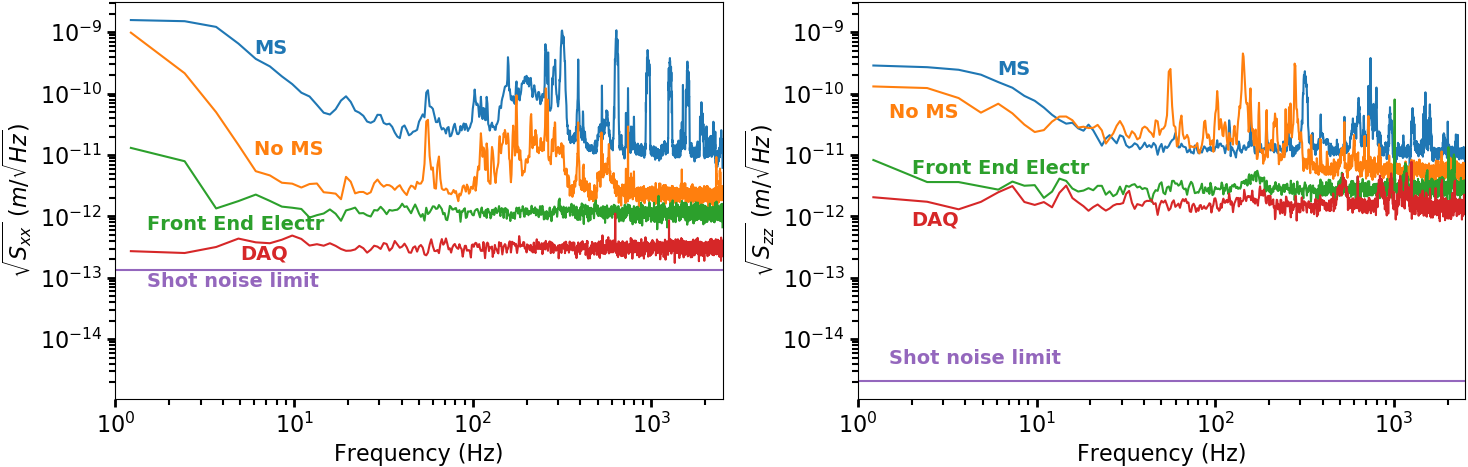}
\caption{(left) Comparison between MS displacement noise in the radial DOF and the shot noise limit.  Also shown is the digitizer noise, with the data-acquisition (DAQ) input terminated, the noise of the photodetector and front-end electronics without incident light, and the noise measured by the full heterodyne readout, but without a MS in the trap. (right) MS displacement noise in the axial DOF. In this case, the data collected without the MS is obtained by reflecting the light off of a gold-plated cantilever at the trap position. The intensity of the trapping beam was tuned such that the cantilever reflected the same power as a MS. The remaining curves are obtained in a manner similar to those in the left panel. Data are calibrated empirically with the spring constant measurement discussed in Sec. III, whereas the radial shot noise calculation makes use of Eq.~(\ref{eq:current_seg_detector}) and the axial shot noise makes use of the interferometric relation, both assuming perfect modes and mode matching.}
\label{fig:noise_compare}
\end{figure*} 

Shot noise places a fundamental limit on the performance of the system, which is computed following Ref.~\cite{Siegman:1966}. The shot-noise-limited displacement spectral density can be determined from the usual relation between shot noise and mean photocurrent, $S_{\text{shot}} = 2 e I$ \cite{Schottky:1918}, together with Eq.~(\ref{eq:current_seg_detector}). We find

\vspace*{-0.2cm}
\begin{equation} \label{eq:radial_shot_noise}
S_{xx} = \frac{\pi e {\mathcal{P}}_{2}}{2 \xi {\mathcal{P}}_{1}} \frac{(w_{1}^{2} + w_{2}^{2})^{3}}{w_{1}^{4}k^2d^2c^2}({\mathcal{P}}_{1} + {\mathcal{P}}_{2}),
\end{equation}

\noindent where $S_{xx}$ refers to displacements along a radial DOF. Analysis of the axial DOF is more straightforward. The axial position of the MS is determined by using heterodyne detection to measure the phase of light reflected by the MS. A change in the phase of the reflected light, $\Delta \phi_z$, is related to axial displacements of the MS, $z_{\text{MS}}$, by  $(\Delta \phi_z / z_{\text{MS}}) = {2 \pi} / (\lambda /2)$. Assuming perfect mode matching, the shot noise limit for the axial position measurement is 

\vspace*{-0.2cm}
\begin{equation} \label{eq:axial_spectral_density}
S_{zz} = \left(\frac{\lambda}{4 \pi}\right)^{2} \frac{e(\mathcal{P}_{A} + \mathcal{P}_{B})}{8 \xi \mathcal{P}_{A} \mathcal{P}_{B} },
\end{equation}

\noindent where $S_{zz}$ refers to displacements along the axial DOF, $\mathcal{P}_{A}$ is the power of the axial reference beam and $\mathcal{P}_{B}$ is the power reflected from the MS. A detailed derivation of Eqs.~(\ref{eq:radial_shot_noise}) and (\ref{eq:axial_spectral_density}) is included in the appendix. Throughout, we assume perfect mode matching, because this represents the fundamental limitation to which any practical implementation should be compared.

The values of $\sqrt{S_{xx}}$ (representative of both radial DOFs) and $\sqrt{S_{zz}}$ are shown in Fig.~\ref{fig:noise_compare}, together with position spectral densities measured under various conditions. The cases shown in the figure correspond to displacement spectra acquired with and without a MS in the trap, with the trapping beam off, and with the front-end electronics disconnected and data-acquisition electronics terminated. The latter two cases measure the photodetector and front-end electronics noise, as well as digitizer noise, respectively.  Clearly, nonfundamental sources of displacement noise far exceed the shot noise limit and thus substantial performance improvements should follow successive refinements of the apparatus.

\section{STRAY LIGHT REJECTION \\ WITH HETERODYNE MEASUREMENTS}

\begin{figure}[b]  
\includegraphics[width=\columnwidth]{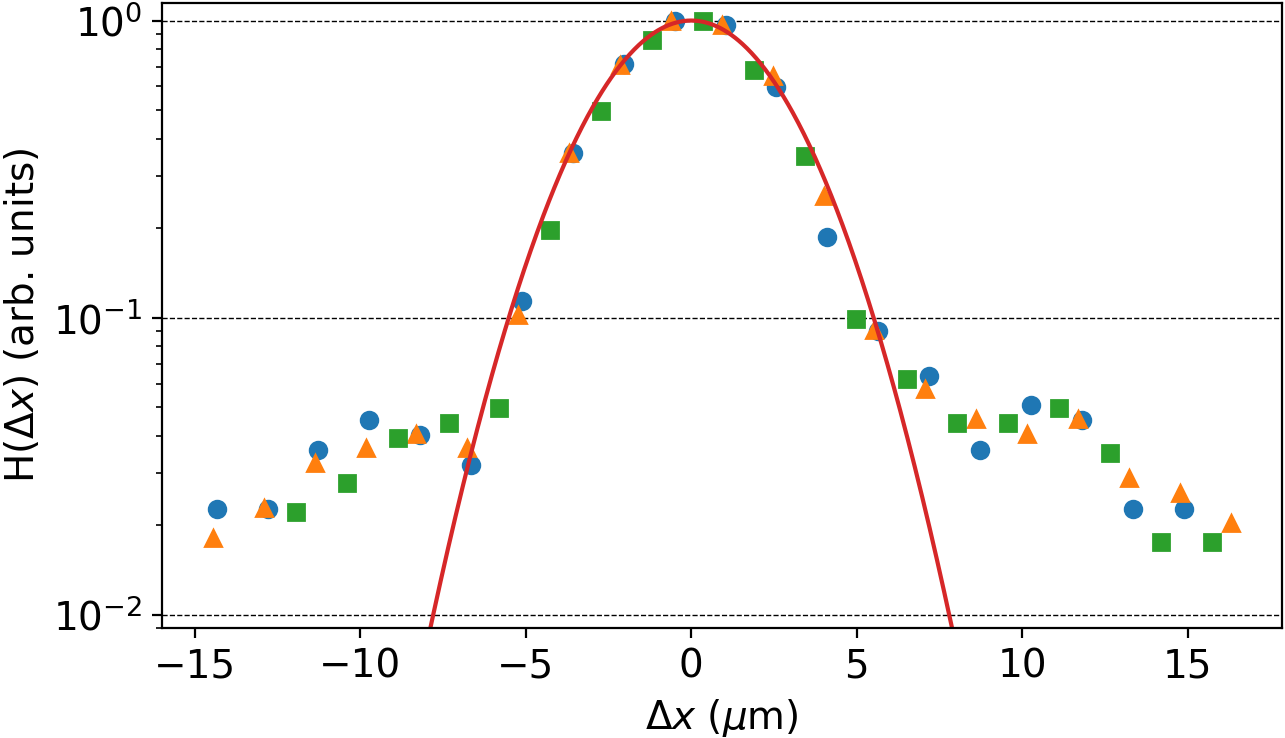}
\caption{Interference contrast vs radial position in the focal plane of the trap, for three distinct sets of measurements, taken for consistency, and shown with differing marker shapes. The solid curve represents the prediction of Eq.~(\ref{eq:ang_rejection}) calculated with the measured beam waists and known focal length. Data is normalized to a maximum of one and centered. The width of the predicted profile is not fit to data. The non-Gaussian tails in the data are likely the result of the halo in the trapping beam, as seen in Fig.~\ref{fig:beam_profile}.}
\label{fig:heterodyne_gauss}
\end{figure} 

Immunity to extraneous sources of light is critical for short-range force sensing where objects that scatter light are brought close to the MS. Heterodyne systems provide substantial rejection of light propagating along a path that is different from the desired one. For a detector positioned in the Fourier plane of the trap, angular rejection corresponds to displacement rejection in the focal plane of the trap. 

The angular rejection of the heterodyne system described is estimated by considering the interference of two Gaussian beams at their focus separated by an angle $\alpha$ between their wavefronts. The profile of angular rejection $H(\alpha)$ can be computed from the normalized integral $ \iint | \vec{E_1} + \vec{E_2} |^2 dA $ with $\vec{E_1}$ and $\vec{E_2}$ the electric fields associated with the appropriately tilted Gaussian beams. We find that the profile of scattered light rejection can be approximated by,

\vspace*{-0.2cm}
\begin{equation} \label{eq:ang_rejection}
H(\Delta x) \simeq \text{exp} \left[ \frac{- (2\pi / \lambda)^2 w_1^2 w_s^2}{4 (w_1^2 + w_s^2)} \left(\frac{\Delta x}{d}\right) ^2 \right],
\end{equation}

\noindent where $w_s$ is the waist associated with the source of scattered light imaged onto the detector, and $\Delta x = d \, \alpha$. This result can be compared with data collected by using the trapping beam as a test source of light and angling the reference beam, with a single-channel photodiode placed in the detector focal plane, in the place of the QPD. The response, calibrated in terms of position at radial distances $\Delta x$ from the center of the trap, is shown in Fig.~\ref{fig:heterodyne_gauss}, along with the prediction of Eq.~(\ref{eq:ang_rejection}). The tails of the distribution present in the data, but not the calculation, are likely due to interference of the reference beam with the halo of the trapping beam, shown in Fig.~\ref{fig:beam_profile}.

\section{ACCELERATION NOISE PERFORMANCE}

\begin{figure}[t]  
\includegraphics[width=\columnwidth]{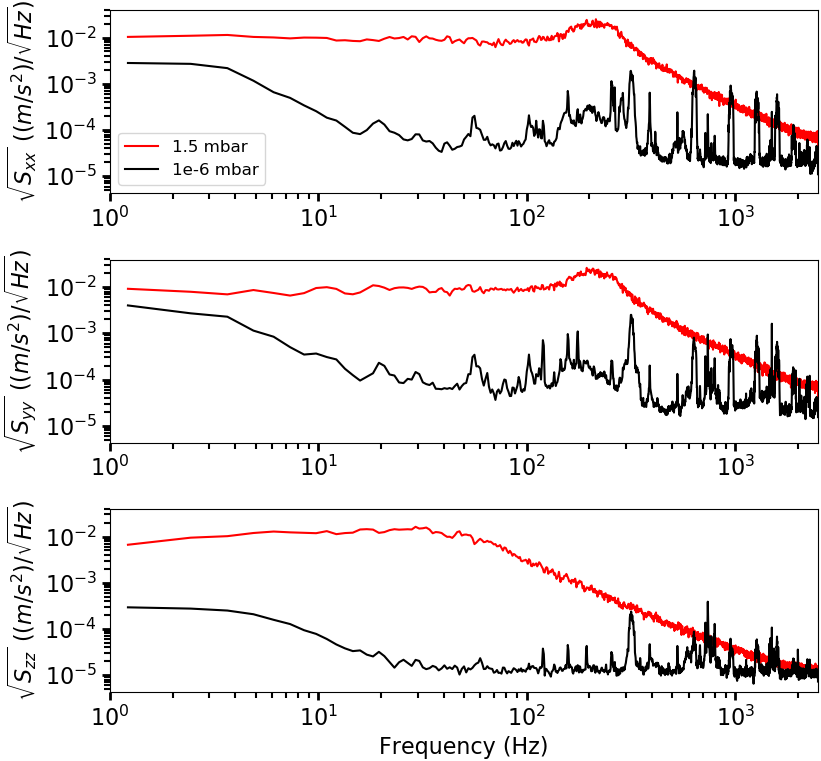}
\caption{Acceleration spectral densities for each of the three DOFs for a MS in the trap at $1.5$ and \SI{e-6}{mbar} of residual gas, with feedback cooling active for the latter case. The data for both axial and radial DOFs were calibrated into physical units following the procedure discussed in Sec. III. The curves for \SI{e-6}{mbar} pressure are directly proportional to the displacement noise with a MS, displayed in Fig.~\ref{fig:noise_compare}.}
\label{fig:bead_spectra}
\end{figure}

Acceleration noise, defined as force noise per unit MS mass, is an important figure of merit for force-sensing applications. While the primary goal of the technique described here is to provide a displacement measurement insensitive to stray sources of light, with comfortable access to the trapping region, the acceleration noise achieved is comparable to the state of the art for levitated MSs. Figure~\ref{fig:bead_spectra} shows the acceleration amplitude spectral densities of the MS motion in each degree of freedom under vacuum conditions ($\SI{e-6}{mbar}$) and, for comparison, at a pressure of \SI{1.5}{mbar} where the MS is driven by collisions with residual gas. The 10 to 100~Hz band where the noise has a broad minimum is used for the force-sensing application of interest to this program. In this band we measure an acceleration noise of $\SI{7.5e-5}{(m/s^{2})/\sqrt{Hz}}$ for the radial DOFs and $\SI{1.5e-5}{(m/s^{2})/\sqrt{Hz}}$ for the axial DOF. 

A comparison of the acceleration noise achieved here with those obtained with other techniques is shown in Table~\ref{tab:noise}. The noise reported in the table corresponds to the optimal conditions reported by the authors, in analogy with the data presented here. Systems optimized for smaller MS are in some cases sensitive to smaller forces, but have poorer acceleration sensitivities [$\gtrsim 0.1\ \SI{}{(m/s^2)/\sqrt{Hz}}$]~\cite{Fonseca:2016,Jain:2016,Vovrosh:2017,Hempston:2017}, and are not included in the table.  All apparatuses capable of trapping MS larger than $\SI{0.1}{\mu m}$, other than the one described here, make use of non-interferometric optical measurements and require auxiliary imaging beams. The acceleration noise observed here is the lowest reported for optically levitated MS.

\begin{table}[b]
  \caption{Comparison of reported radial-displacement and \protect\linebreak -acceleration noise, in a variety of optical trapping apparatuses using MSs with diameters $> \SI{0.3}{\mu m}$. Cases designed for smaller MS are not included since they are not optimized for acceleration sensitivity. The figures reported in \cite{Li:2011} are extracted from the case with a MS in their trap, in order to provide a valid comparison. Shown are: the MS radius R, the trap frequency $f = \Omega/(2\pi)$, the displacement noise $\sigma_x$, and acceleration noise $\sigma_a$ in a frequency band $(f_1,f_2)$. The radial DOFs are chosen here, because they are more relevant for force-sensing programs.}
    \label{tab:noise}
    \begin{ruledtabular}
    \begin{tabular}{lllccc}
	Ref.&    \multicolumn{1}{c}{$R$}    & \multicolumn{1}{c}{$f$} & $\sigma_x$ & $\sigma_a$ & ($f_1,f_2$) \\
    & ($\SI{}{\mu m}$) & ($\SI{}{kHz}$) & ($\SI{}{m/\sqrt{Hz}}$) & [$\SI{}{(m/s^2)/\sqrt{Hz}}$] &       (kHz)        \\
    \hline
    \midrule
    \vspace{-0.2cm} \\
    This work         &  2.4  &  0.25   &  \SI{3.1e-11}{}   &  \SI{7.5e-5}{}  &   (0.01,0.1)    \\
    \cite{Li:2011}       &   1.5   &    9.1    &  \SI{1.4e-13}{}   &  \SI{4.6e-4}{}  &     (1,10) \\
    \cite{Ranjit:2015}   &  1.5  &  1.0    &  \SI{2.0e-10}{}   &  \SI{7.7e-3}{}  &      (0.01,1) \\
    \cite{Ranjit:2016}   &  0.15  &  2.8    &  \SI{1.8e-10}{}   &  \SI{5.7e-2}{}  &      (0.01,3) \\
    \bottomrule
    \end{tabular}
    \end{ruledtabular}
\end{table}


\section{CONCLUSIONS}

We have described a technique applying heterodyne detection to measure the three-dimensional position of a microsphere in an optical trap.  This technique allows all functions (trapping, feedback, and position measurement) to be performed with a single laser, while providing a substantial rejection of signals arising from scattered light. This provides unmatched access to the trapped microsphere and, because of the insensitivity to scattered light, is particularly powerful in applications where the microsphere is used as a force sensor in close proximity to other objects.  

We have presented the current performance of the system in terms of scattered light rejection and noise, which are at the state-of-the-art level. The noise performance of the system is far from the fundamental limit imposed by shot noise, leaving significant room for improvement.

Our group is planning to apply this technique to the measurement of interactions at sub-\SI{100}{\mu m} distance that may arise from non-Newtonian gravity.

\section*{ACKNOWLEDGMENTS}

We would like to thank J.~Fox (SLAC), L.~Holberg (Stanford), and R.~Adhikari (Caltech) for useful discussions during the early stages of this work, T.~Morris and K.~Urbanek (Stanford) for their guidance with the fiber-optic system as well as R.~DeVoe and A.~Kawasaki (Stanford) for their comments on earlier drafts of this manuscript.  This work was supported, in part, by NSF Grant No. PHY1502156 and by the Heising-Simons foundation. A.D.R. is supported by an ARCS Foundation Stanford Graduate Fellowship.

\vspace*{0.2cm}


\section*{APPENDIX: CALCULATIONS}

\subsection*{1. Radial-displacement calibration}

Here we provide a detailed derivation of Eq.~(\ref{eq:current_seg_detector}), which is the difference in photocurrent at the heterodyne frequency between adjacent sides of a segmented detector per radial displacement of the MS. To begin, following Sec. III, if a MS deflects a trapping beam of power $\mathcal{P}$ by an angle $\theta$, the restoring force from the change in optical momentum flux is given by,

\vspace*{-0.2cm}
\begin{equation} \label{eq:force_to_displacement_app}
F_{\text{opt}}= \frac{\mathcal{P}}{c} \text{sin}[\theta] \approx \frac{\mathcal{P}}{c} \theta \approx \frac{\mathcal{P}}{c} \frac{x_B}{d}, \tag{A1}
\end{equation}

\noindent where $c$ is the speed of light and the approximation assumes small deflections. We have related optical force to displacements of the transmitted beam, $x_B$, since $x_B \approx d \theta$ for small $\theta$, where $d$ is the focal length of the recollimation lens.

Consider the interference of two Gaussian beams on a segmented detector, where both beams have foci in the detector plane, and one beam is displaced by a radial distance $x_B$. Let the segments be half-infinite planes with their border parallel to the $y$ axis of our detector at $x=0$. Let the center of our reference beam be at the origin and the trapping beam displaced by $\vec{\Delta x} = x_B \hat{x}$.

The electric field of a Guassian beam at its focus is, 

\vspace*{-0.2cm}
\begin{align}
\vec{\widetilde{E}} (\vec{x}, t) &= \vec{E}(\vec{x}) e^{i \phi} \text{exp} \left[ i \omega t \right] \notag \\
&= E_o \hat{p} \, \text{exp} \left[ \frac{-|\vec{x}|^2}{w_0^2} \right] e^{i \phi} \text{exp} \left[ i \omega t \right], \tag{A2}
\end{align}

\noindent where $E_o$ is the peak electric field, $\phi$ is a phase, $\hat{p}$ is the polarization, $w_o$ is the waist, and $\omega$ is the optical angular frequency. If we interfere two such beams that differ in frequency by an amount $\Delta \omega$, the resulting irradiance on a photodetector can be computed as the square of the sum of the electric fields. Dropping the constant terms and considering the term oscillating at $\Delta \omega$, we found,

\vspace*{-0.2cm}
\begin{equation} \label{eq:photocurrent_app}
\hspace*{-0.5cm} \mathclap{I_{IF}\! =\! \frac{2 \xi}{\eta}\! \iint\! \vec{E_1}(\vec{x}_1\!)\! \cdot\! \vec{E_2}(\vec{x}_2\!) \text{cos} \! \left[ \Delta \omega \, t \! + \! \Delta \phi \right] dA_1,} \tag{A3}
\end{equation}
\vspace*{0.1cm}

\noindent where $\xi$ is the responsivity of the photodetector in [\SI{}{A/W}], $\eta = 1 / c \, \epsilon$ is the wave impedance, $\epsilon$ is the dielectric constant, and the integral is computed over some bounded segment. $\vec{E}_i$ are the electric fields in the detector plane, with $i = 1$ the reference beam and $i = 2$ the trapping beam, and $\Delta \phi$ is a common-mode phase across all quadrants due to path length fluctuations. The expression is defined in terms of and integrated over $\vec{x}_1$ with $\vec{x}_2 = \vec{x}_1 + \vec{\Delta x}$.

We perform this integral over two regions: (1) $\vec{x}_1 \cdot \hat{x} \in (-\infty, 0]$, $\vec{x}_1 \cdot \hat{y} \in (-\infty, \infty)$ and (2) $\vec{x}_1 \cdot \hat{x} \in [0, \infty)$, $\vec{x}_1 \cdot \hat{y} \in (-\infty, \infty)$. Finally, we take the difference of these two integrals to find an expression for $\Delta I$ in terms of beam displacement $x_B$. The necessary integral is given by,

\begin{widetext}
\vspace*{-0.2cm}
\begin{align}
\int_{-\infty}^0 dx \int_{-\infty}^{+\infty} dy \, e^{ - a (x^2 + y^2) / w_1^2 - a ((x+ x_B)^2 + y^2) / w_2^2} &= \frac{\pi w_1^2 w_2^2 e^{-a x_B^2 / (w_1^2 + w_2^2)}}{2 \, a (w_1^2 + w_2^2)}  \left[ 1 + \text{Erf} \left( \frac{\sqrt{a} x_B}{w_2^2} \sqrt{\frac{w_2^2 w_1^2}{w_1^2 + w_2^2}} \right) \right]  \notag \\
&\approx \frac{\pi w_1^2 w_2^2}{2 \, a (w_1^2 + w_2^2)} \left( 1 + \frac{2}{\sqrt{\pi}} \cdot \frac{\sqrt{a} x_B}{w_2} \sqrt{\frac{w_1^2}{w_1^2 + w_2^2}} + .\,.\,. \right), \tag{A4}
\end{align}
\end{widetext}


\noindent where $a$ is a constant and $a=2$ for the integrals performed here. The result has been expanded to linear order in $x_B$, because we expect $x_B / w_i \ll 1$. The result of this calculation is,

\vspace*{-0.2cm}
\begin{equation} \label{eq:current_seg_detector_outbeam}
\frac{\Delta I}{x_B} = 4 \xi \sqrt{\frac{\mathcal{P}_1 {\mathcal{P}}_2}{4 \pi}} \frac{w_1^2 }{( w_1^2 + w_2^2)^{3/2}} \text{cos} [ \Delta \omega t + \Delta \phi], \tag{A5}
\end{equation}

\noindent where we have related the peak electric field in a Gaussian beam to its total power by integrating the irradiance of a single, ideal Gaussian beam which yields $\mathcal{P} = \frac{1}{4}\pi c \epsilon E_o^2 w_o^2$.

Finally, we can find Eq.~(\ref{eq:current_seg_detector}) by using Eqs.~(\ref{eq:force_to_displacement_app}) and (\ref{eq:current_seg_detector_outbeam}) and the harmonic-oscillator assumption that $F_{\text{opt}} = k x_{\text{MS}}$ with $k$ being the trap spring constant,

\vspace*{-0.2cm}
\begin{equation} \label{eq:current_seg_detector_final}
\frac{\Delta I}{x_{\text{MS}}}\! =\! \frac{\Delta I}{x_B} \left( \frac{x_B}{x_{\text{MS}}} \right) \! =\! \frac{\Delta I}{x_B} \left( \frac{\frac{F_{\text{opt}} c d}{\mathcal{P}_2}}{\frac{F_{\text{opt}}}{k}} \right)\! =\! \frac{\Delta I}{x_B} \cdot \frac{kcd}{\mathcal{P}_2}, \tag{A6}
\end{equation}

\noindent which yields the result quoted in Sec.~III.

\subsection*{2. Radial-displacement noise}

In this section we compute the radial-displacement noise quoted in Sec.~IV, which is a far more simple calculation. We assume that the radial displacement noise is simply the photocurrent shot noise multiplied by the square of the radial-displacement calibration computed previously.

The photocurrent shot noise is given by Schottky's result, $S_{shot} = 2 e I$ \cite{Schottky:1918}, with $e$ being the fundamental charge and $I$ the mean photocurrent. Computing directly,

\vspace*{-0.2cm}
\begin{align}
S_{xx} &= S_{shot} \cdot \left( \frac{x_{\text{MS}}}{\Delta I} \right)^2 \notag \\
&= (2 e I) \cdot \left( \sqrt{ \frac{4 \pi}{ \mathcal{P}_1 \mathcal{P}_2} } \frac{\mathcal{P}_2}{4 \, \xi \, k \, d \, c} \frac{(w_1^2 + w_2^2)^{3/2}}{w_1^2} \right)^2 \notag \\
&= (2 e \xi (\mathcal{P}_1 + \mathcal{P}_2 )) \cdot \left( \frac{\pi \mathcal{P}_2}{4 \,\mathcal{P}_1 \, (\xi k d c)^2} \frac{(w_1^2 + w_2^2)^3}{w_1^4} \right) \notag \\
&= \frac{\pi e \mathcal{P}_2}{2 \xi \mathcal{P}_1} \frac{(w_1^2 + w_2^2)^3}{w_1^4 k^2 d^2 c^2} (\mathcal{P}_1 + \mathcal{P}_2) \tag{A7}
\end{align}

\noindent where we have dropped the portion of $(\Delta I / x_{\text{MS}})$ that oscillates in time, since we measure the amplitude of the interference. This is exactly the result quoted in Sec.~IV.

\subsection*{3. Axial-displacement noise}

Our expression for axial-displacement noise is derived from error propagation. The axial signal is determined by comparing the ratio of neighboring samples of the interference signal generated by light reflected from the MS. Because $f_{ADC} = 4 \cdot f_{IF}$, with $f_{ADC}$ being the sampling frequency and $f_{IF}$ in the interference frequency, the arc tangent of this ratio can be interpreted as the phase, $\phi$, of the signal and can be averaged over many cycles of the interference.

Let the $z$ signal be given by,

\vspace*{-0.2cm}
\begin{equation} \label{eq:z_signal}
z_{\text{MS}} = \frac{\lambda / 2}{2 \pi} \text{atan} \left[ \frac{A_i}{A_{i\pm1}} \right], \tag{A8}
\end{equation}

\noindent where $\lambda$ is the wavelength of light and the prefactor relates path-length changes, due to MS motion, to the phase of the back reflection, and thus the phase of the interference signal. $A_i$ are neighboring voltage samples with shot noise spectral density $S_{A_i} = G_z^2 R_{t,z}^2 (2 e I)$, with $G_z$ being the $z$ electronics gain and $R_{t,z}$ the $z$ transimpedance. We assume the variance of the signal $z_{\text{MS}}$ can be given directly by propagating errors,

\begin{widetext}
\vspace*{-0.2cm}
\begin{align} \label{eq:z_noise}
S_{zz} &= \left| \frac{\partial z_{\text{MS}}}{\partial A_i} \right|^2 S_{A_i} + \left| \frac{\partial z_{\text{MS}}}{\partial A_{i\pm1}} \right|^2 S_{A_{i\pm1}}, \notag \\
&= \left( \frac{\lambda / 2}{2 \pi} \right)^2 2 G_z^2 R_{t,z}^2 e I \Bigg[ \left(\frac{1}{\left(1+\frac{A_i^2}{A_{i\pm1}^2}\right) A_{i\pm1}}\right)^2  + \left(\frac{A_i}{\left(1+\frac{A_i^2}{A_{i\pm1}^2}\right) A_{i\pm1}^2}\right)^2 \Bigg] \notag \\
&= \left( \frac{\lambda}{4 \pi} \right)^2 \frac{2 G_z^2 R_{t,z}^2 e \xi (\mathcal{P}_A + \mathcal{P}_B)}{A_i^2 + A_{i\pm1}^2}, \tag{A9}
\end{align}
\end{widetext}


\noindent where $\mathcal{P}_A$ and $\mathcal{P}_B$ are the power of the back-reflected and reference beams, respectively, although the result is symmetric with regard to these powers.

The signal samples $A_i$ are voltage amplitudes of the interference signal, sampled every quarter wavelength, and can thus be expressed as,

\vspace*{-0.2cm}
\begin{align}
A_i = 4 G_z R_{t,z} \xi \sqrt{\mathcal{P}_A \mathcal{P}_B} \begin{cases} \text{sin}(\phi) & i=1,5,.\,.\,. \\ -\text{cos}(\phi) & i = 2,6,.\,.\,. \\ -\text{sin}(\phi) & i=3,7,.\,.\,. \\ \text{cos}(\phi) & i=4,8,.\,.\,.\,, \end{cases} \tag{A10}
\end{align}

\noindent which assumes perfect mode matching. The coefficient can be derived from Eq.~\ref{eq:photocurrent_app} without displacing one of the beams, using identical waists and converting current to voltage. Substituting these expressions and noting sin$^2$+cos$^2$=1, 

\vspace*{-0.2cm}
\begin{align} \label{eq:z_noise_final}
S_{zz} &=  \left(\frac{\lambda}{4 \pi}\right)^{2} \frac{e(\mathcal{P}_{A} + \mathcal{P}_{B})}{8 \xi \mathcal{P}_{A} \mathcal{P}_{B} }, \tag{A11}
\end{align}

\noindent we immediately find the result quoted in Sec.~IV.

\bibliographystyle{apsrev4-1}
\bibliography{interferometric_optical_levitation_v2.0}

\end{document}